\documentclass[letterpaper, 10 pt, conference]{ieeeconf}  

\IEEEoverridecommandlockouts                              

\usepackage{graphicx}
\usepackage{url}  

\usepackage{amsmath}
\DeclareMathOperator*{\argmax}{argmax}

\title{\LARGE \bf
Inception-Based Network and Multi-Spectrogram Ensemble \\ Applied For Predicting Respiratory Anomalies and Lung Diseases
}

\author{Lam~Pham$^{1}$,
             Huy~Phan$^{2}$, 
             Ross~King$^{1}$,
             Alfred~Mertins$^{3}$,
             Ian~McLoughlin$^{4}$ %
\thanks{L. Pham and R. King are with Competence Unit Data Science \& Artificial Intelligence, Center for Digital Safety \& Security, Austria Institute of Technology, Austria.}%
\thanks{H. Phan is with School of Electronic Engineering and Computer Science, Queen Mary University of London, UK.}%
\thanks{A. Mertins is with the Institute for Signal Processing, University of L\"ubeck, Germany.}%
\thanks{I. McLoughlin is with Singapore Institute of Technology, Singapore.}%
}

\begin{document}

\maketitle
\thispagestyle{empty}
\pagestyle{empty}

\begin{abstract}

This paper presents an inception-based deep neural network for detecting lung diseases using respiratory sound input.
Recordings of respiratory sound collected from patients are firstly transformed into spectrograms where both spectral and temporal information are well presented, referred to as front-end feature extraction.
These spectrograms are then fed into the proposed network, referred to as back-end classification, for detecting whether patients suffer from lung-relevant diseases.
Our experiments, conducted over the ICBHI benchmark meta-dataset of respiratory sound, achieve competitive ICBHI scores of 0.53/0.45 and 0.87/0.85 regarding respiratory anomaly and disease detection, respectively.
\newline

\indent \textit{Clinical relevance}--- Respiratory disease, wheeze, crackle, inception, convolutional neural network. 
\end{abstract}
\section{INTRODUCTION}

The World Health Organization informed that one of the most common mortality factors worldwide is the respiratory illness~\cite{who}. 
To combat morality from respiratory diseases, the most effective way is early detection that not only helps to limit spread but also improves treatment effectiveness. 
During a lung auscultation which is an important aspect of a medical examination, experts can hear, detect anomaly sounds such as \textit{Crackles} or \textit{Wheezes}, thus diagnose respiratory-relevant diseases.
Therefore, if these anomaly sounds can be automatically detected by an edge device, it is very useful for self-observation, and thus early detect the diseases.
Analysing respiratory sound was early mentioned in ~\cite{sound_early_03, sound_early_01, sound_early_02}, and recently this research topic has attracted a lot of attention and several machine learning methods have been proposed.
In particular, frame-based represented systems proposed in~\cite{lung_hmm_01} and \cite{lung_hmm_02} applied Mel-frequency cepstral coefficient (MFCC) extraction, a robust feature popularly used in Automatic Speech Recognition (ASR), to derive feature likely vectors.
Next, these vectors are explored by conventional machine learning methods such as Hidden Markov Model~\cite{lung_hmm_01, lung_hmm_02}, Support Vector Machine~\cite{lung_svm_01}, or Decision Tree~\cite{lung_tree_18}.
Meanwhile, approaches relying on spectrogram representation involves generating two-dimensional spectrogram (i.e. an image), which is then fed into powerful network architectures such as CNN~\cite{lung_cnn_01, lung_cnn_02} or RNN~\cite{lung_rnn_01, lung_rnn_02} for classification.
Although recent publications which applied machine learning techniques report good performance, it is hard to compare among systems due to different training/test data ratios used as well as various experiments conducted over proprietary datasets. To make our work comparable, we evaluate our systems on the 2017 Internal Conference on Biomedical Health Informatics (ICBHI)~\cite{lung_dataset}, which is one of the largest public benchmark respiratory sound dataset. Furthermore, we obey the ICBHI challenge setup, thus use the ratio of 60/40 for training/test sets defined by the challenge~\cite{icb_ratio}, in which a subject is not presented in both training and test sets (note that some systems randomly separate ICBHI dataset into training and test subsets regardless this patient independency~\cite{lung_cnn_01, lung_cnn_02, lung_rnn_01, lam02}). 
Regarding our proposed system, we firstly apply Wavelet and Gammatone transformation to generate Scalogram and Gammatonegram from an audio signal, respectively. These spectrograms are then fed into the proposed inception-based deep neural network to detect respiratory anomalies and lung diseases.

\section{ICBHI dataset and tasks defined}
The ICBHI dataset~\cite{lung_dataset}, which was collected from a total of 128 patients over 5.5 hours, comprises 920 audio recordings with a wide range of sampling frequencies ranging from 4 to 44.1\,kHz and various lengths from 10 to 90\,s. In each recording, four different types of cycles (\textit{Crackle}, \textit{Wheeze}, \textit{Both Crackle \& Wheeze}, and \textit{Normal}) are marked with onset and offset time. Additionally, each recording is also associated to identifies the patient's disease status, mainly classified into three main categories: \textit{Chronic Disease} (i.e. COPD, Bronchiectasis and Asthma), \textit{Non-Chronic Disease} (i.e. Upper and Lower respiratory tract infection, Pneumonia, and Bronchiolitis), and \textit{Healthy}.
Given ICBHI metadata, this paper proposes two main tasks. Firstly, Task 1 aims to classify four different types of respiratory cycles mentioned. Secondly, Task 2, referred to as respiratory disease prediction, is to detect whether a patient suffers from \textit{Chronic Diseases}, \textit{Non-Chronic Diseases}, or \textit{Healthy}.
Regarding experimental setting, we obey ICBHI challenge, split the audio recordings into training/test sets with a ratio of 60/40 without recordings of the same subject presenting in both training and test data. While entire audio recordings are used to evaluated in Task 2, respiratory cycles with onset and offset labels are extracted from the entire recordings for experiments conducted in Task 1. To evaluate performance and compared with the state-of-the-art systems, we  use metrics of  Sensitivity (Sen.), Specitivity (Spec.) and ICBHI scores (Average score (AS) and Harmonic score (HS)) that are also the ICBHI criteria as mentioned in ~\cite{pham2020_01, ic_cnn_19_iccas, pham2020cnnmoe}.

\section{Proposed baseline system}
\label{baseline}
\subsection{The baseline system architecture}
%
\begin{figure}[h]
    	\vspace{-0.2cm}
    \centering
    \includegraphics[width =1.0\linewidth]{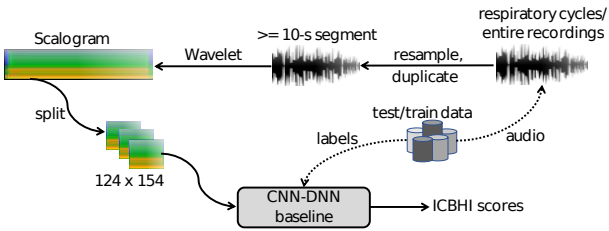}
    	\vspace{-0.5cm}
	\caption{The baseline system architecture.}
    \label{fig:A1}
            	\vspace{-0.3cm}
\end{figure}
\begin{table}[h]
    \caption{The baseline network architecture proposed} 
        	\vspace{-0.2cm}
    \centering
    \scalebox{0.9}{
    \begin{tabular}{l c} 
        \hline 
            \textbf{CNN-DNN baseline}   &  \textbf{Output}  \\
               \textbf{architecture}  &  \textbf{($W{\times}H{\times}C$)}  \\

        \hline 
         \textbf{CNN} & \\
         Input layer (image patch of $124{\times}154{\times}1$)  &        \\
         Bn - Cv [$3{\times}3$] - Relu - Bn - Mp [$2{\times}2$] - Dr ($10\%$)      & $62{\times}78{\times}64$\\
         Bn - Cv [$3{\times}3$] - Relu - Bn - Mp [$2{\times}2$] - Dr ($15\%$)      & $31{\times}39{\times}128$\\
         Bn - Cv [$3{\times}3$] - Relu - Bn - Mp [$2{\times}2$] - Dr ($20\%$)      & $16{\times}20{\times}256$\\
         Bn - Cv [$3{\times}3$] - Relu - Bn - Gmp - Dr ($25\%$)       & $512$\\
         \hline
         \textbf{DNN} & \\
         Input layer ($512$-dimensional vectors)  &        \\
         Fl - Relu - Dr ($30\%$) &  $1024$       \\
         Fl - Softmax & C \\
       \hline 
    \end{tabular}
    }
    \label{table:CDNN} 
\end{table}
To evaluate the proposed system, we use a baseline system as shown in Fig. \ref{fig:A1} for comparison.
In particular, we re-sample respiratory cycles in Task 1 to 4\,Khz as frequency bands of abnormal sounds (\textit{Crackle} and \textit{Wheeze}) locate around 60-2000 Hz~\cite{lung_rnn_01}. For entire recordings in Task 2, we resample them to 16\,Khz to overcome different recording sample rates.
Re-sampled respiratory cycles or entire recordings showing different lengths are next duplicated to ensure the same length of 10 seconds for respiratory cycles in Task 1 and minimum 10 seconds in Task 2, respectively.
Next, respiratory cycles go through a band-pass filter of 100-2000 Hz to reduce noise (note that band-pass filtering is not applied to entire recordings in Task 2). After that, these respiratory sounds are transformed into Scalogram by using continuous Wavelet transformation with \textit{Morse} as Wavelet mother function.
Each 10-s Scalogram of one respiratory cycle in Task 1 is thus scaled into a size of $124{\times}154$ image.
Although the same scale ratio is also applied, the Scalograms of entire recordings in Task 2 show various time resolutions as original recordings' lengths are different (note that the frequency resolution of 124 is same for two tasks).
Therefore, the long Scalograms of entire recordings in Task 2 are separated into various non-overlapped image patches of $124{\times}154$ that has the same size as 10-s Scalograms in Task 1.
To enlarge Fisher’s criterion (i.e. the ratio of the between-class distance to the within-class variance in the feature space), we apply mixup data augmentation~\cite{mixup2} over image patches of $124{\times}154$ to increase variation of the training data.

For back-end classification, we propose a CNN-DNN baseline network architecture shown in Table \ref{table:CDNN}.
In particular, the baseline architecture is separated into two main parts: CNN and DNN.
The CNN contains sub-blocks which perform batch normalization (Bn), convolution (Cv[kernel number, kernel size]), rectified linear units (Relu), max pooling (Mp[kernel size]), global max pooling (Gmp), dropout (Dr (percentage drop)), configured as shown in the upper part of Table \ref{table:CDNN}.
The DNN part contains two fully-connected layers (Fl). While the first fully-connected layer is  followed by rectified linear units (Relu) and dropout (Dr (percentage drop)), Softmax is used after the second fully-connected layer to create probability among categories classified. $C$ takes values of 4 or 3 depending on the number of categories in Task 1 or Task 2, respectively.
\subsection{Experimental setting for the baseline}
As mixup data augmentation is used, labels are not one-hot encoding format. Therefore, we use Kullback–Leibler divergence (KL) loss shown in Eq. (\ref{eq:kl_loss}) below.
\begin{align}
   \label{eq:kl_loss}
   Loss_{KL}(\Theta) = \sum_{n=1}^{N}\mathbf{y}_{n}\log \left\{ \frac{\mathbf{y}_{n}}{\mathbf{\hat{y}}_{n}} \right\}  +  \frac{\lambda}{2}||\Theta||_{2}^{2}
\end{align}
where  \(\Theta\) are trainable parameters, constant \(\lambda\) is initially set to $0.0001$, $N$ is batch size set to 100, $\mathbf{y_{i}}$ and $\mathbf{\hat{y}_{i}}$  denote expected and predicted results, respectively.
We construct the proposed baseline with Tensorflow framework and the training is carried out for 100 epochs using Adam~\cite{Adam} for optimization.

As Task 2 evaluates over entire recordings while the proposed CNN-DNN baseline network works on one image patch of $124{\times}154$, the result over an entire recording is obtained by averaging results over its patches. 
Let consider $\mathbf{p}^{m} = (p_{1}^{m}, p_{2}^{m},\ldots,p_{C}^{m})$ as the probability obtained from the \(m^{th}\) out of \(M\) patches and $C$ is the number of categories classified. Then, the mean probability of an entire recording instance is denoted as \(\mathbf{\bar{p}} = (\bar{p}_{1}, \bar{p}_{2},\ldots, \bar{{p}}_{C})\) where
\begin{equation}
\label{eq:mean_stratergy_patch}
\bar{p}_{c} = \frac{1}{M}\sum_{m=1}^{M}p_{c}^{m}  ~~~  \text{for}  ~~ 1 \leq c \leq C
\end{equation}
The predicted label  \(\hat{y}\) is then determined as 
\begin{equation}
\label{eq:average}
\hat{y} = \argmax_{c \in \{1,2,\ldots,C\}}\bar{p}_c.
\end{equation}

\section{An analysis of inception-based network architecture and ensemble of multiple spectrogram input}
Given by the proposed baseline, we evaluate whether the proposed inception-based network architectures and ensemble of different spectrograms are useful to improve the performance. 

\subsection{Inception-based deep neural network}
\begin{figure}[h]
    \centering
    \includegraphics[width =0.9\linewidth]{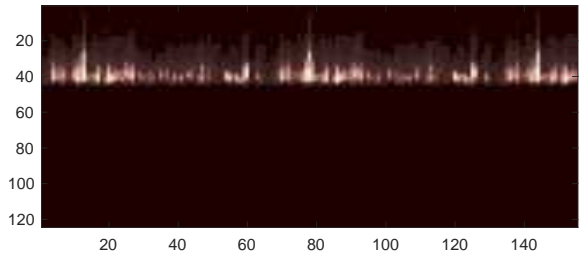}
    	\vspace{-0.2cm}
	\caption{Scalogram of Wheeze cycles}
    \label{fig:A2}
        	\vspace{-0.3cm}
\end{figure}
\begin{figure}[h]
    \centering
    \includegraphics[width =0.95\linewidth]{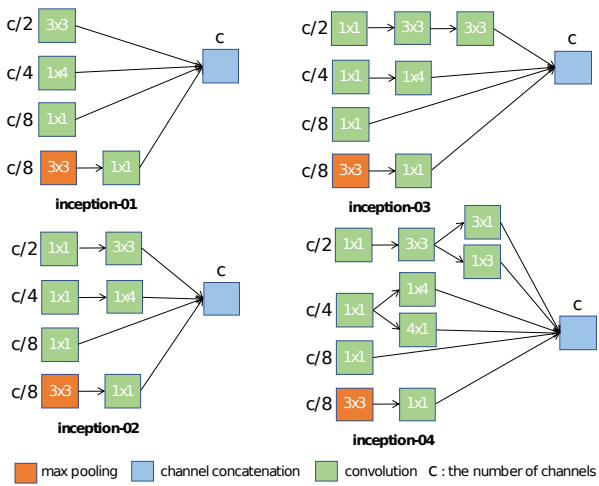}
    	\vspace{-0.3cm}
	\caption{Inception layer architectures.}
    \label{fig:A3}
            	\vspace{-0.3cm}
\end{figure}

Look at a 10-s Scalogram of a Wheeze cycle represented as an image with size of $124{\times}154$ as shown in Fig. \ref{fig:A2} (note that the short-time Wheeze cycle is duplicated for three times to obtain 10-s duration in Fig. \ref{fig:A2}), Wheeze spectrum locates within a narrow frequency band (i.e. the narrow band indices from 25 to 40 of 124 central frequencies distributed from minimum frequency of 100 Hz and maximum frequency of 2000 Hz) and shows short time duration (note that Crackle cycles also locate narrow bands). 
To enforce the back-end classification model to learn these minor variation of spatial features, inception-based networks, which perform well on image data~\cite{inception_01}, are applied in this paper.
In particular, we replace convolutional layers used in the CNN part of CNN-DNN baseline by different inception layer architectures as shown in Fig. \ref{fig:A3}. Notably, we use kernel [$1\times4$] instead of  [$5\times5$] as usual to enforce the network focus on minor variation across frequency dimension of spectrum of Wheeze and Crackle sounds. 

\subsection{Ensemble of multiple spectrogram input}
Inspire by~\cite{ic_cnn_19_iccas} that shows ensemble models of different spectrograms beneficial to improve the performance, we evaluate the combination of two Scalogram (two Scal. for short) generated from two different Wavelet mother functions (\textit{Morse} and \textit{Amor}). 
We also evaluate another combination between Scalogram (using \textit{Morse} function) and Gammatonegram (using Gammatone filter~\cite{aud_tool}).
Regarding Gammatone (Gam.) spectrogram, we use the setting of window size = 512, hop size = 256, and filter number = 124 to generate the same patch size of $124{\times}154$ as the Scalogram (Scal.) mentioned in the baseline system in Section \ref{baseline}.
Meanwhile, back-end classifier is reused from the baseline system proposed.
To ensemble two baseline models each of which learns from one type of input, we fuse the probabilities as Eq. (\ref{eq:mean_stratergy_patch})
\begin{equation}
\label{eq:mean_stratergy_patch}
\bar{\mathbf{p}} = \frac{1}{K}\sum_{k=1}^{K} \mathbf{p}^{k}  
\end{equation}
where $\mathbf{p}^{k}$ is the probability outputs obtained from spectrogram $k$ and $\bar{\mathbf{p}}$  is the probability outputs averaged over $K$ spectrograms. Eventually, the final result is obtained by applying likelihood maximization in Eq. (\ref{eq:average})
\section{Experiments and results}
\subsection{Affect of inception-based network architectures}
\begin{table}[h]
	\caption{Respiratory anomaly detection - Task 1} 
        	\vspace{-0.2cm}
    \centering
    \scalebox{0.9}{
    \begin{tabular}{c c c c c c c} 
        \hline 
	    \textbf{Task}  &\textbf{Systems}    &\textbf{Spec.}   &\textbf{Sen.}   &\textbf{AS/HS Scores}  \\
        \hline 
	    Task 1                 &Baseline            &0.68             &0.30            &0.49/0.42  \\
	    Task 1                 &Inception-01     &0.73             &0.30            &\textbf{0.52/0.43}  \\
	    Task 1                 &Inception-02     &0.70             &0.30            &0.50/0.42  \\
	    Task 1                 &Inception-03     &0.69             &0.33            &0.51/0.44  \\
    	    Task 1                 &Inception-04     &0.70             &0.32            &0.51/0.44  \\
       \hline 
    \end{tabular}
    }
    \label{table:cyc_inc} 
\end{table}
\vspace{-0.2cm}
\begin{table}[h]
	\caption{Respiratory disease detection - Task 2} 
        	\vspace{-0.2cm}
    \centering
    \scalebox{0.9}{
    \begin{tabular}{c c c c c c c} 
        \hline 
	    \textbf{Task}  &\textbf{Systems}    &\textbf{Spec.}   &\textbf{Sen.}   &\textbf{AS/HS Scores}  \\
        \hline 
	    Task 2                 &Baseline            &0.59             &0.75            &0.67/0.66  \\
	    Task 2                 &Inception-01     &0.88             &0.81            &0.85/0.84  \\
	    Task 2                 &Inception-02     &\textbf{1.00}             &0.75            &\textbf{0.87/0.85}  \\
	    Task 2                 &Inception-03     &0.53             &\textbf{0.83}            &0.68/0.64  \\
    	    Task 2                 &Inception-04     &0.47             &0.81            &0.64/0.59  \\
       \hline 
    \end{tabular}
    }
    \label{table:ent_inc} 
\end{table}
\vspace{-0.2cm}

As shown in Tables \ref{table:cyc_inc} and \ref{table:ent_inc}, the proposed inception-based networks outperform the baseline over both the classification tasks. In Task 1, \textit{inception-01} network achieves the best scores of 0.52/0.43. Meanwhile, Task 2 of lung disease detection shows the best scores of 0.87/0.85 obtained by \textit{inception-02} architecture.

\subsection{Affect of multiple-spectrogram ensemble}

\begin{table}[h]
	\caption{Respiratory anomaly detection - Task 1} 
        	\vspace{-0.2cm}
    \centering
    \scalebox{0.9}{
    \begin{tabular}{c c c c c c c} 
        \hline 
	    \textbf{Task}  &\textbf{Systems}    &\textbf{Spec.}   &\textbf{Sen.}   &\textbf{AS/HS Scores}  \\
        \hline 
	    Task 1                 &Baseline            &0.68             &0.30            &0.49/0.42  \\
	    Task 1                 &Two Scal.          &0.73             &0.29            &0.51/0.41  \\
	    Task 1                 &Gam. \& Scal.      &0.72            &0.31             &\textbf{0.51/0.43}  \\
        
       \hline 
    \end{tabular}
    }
    \label{table:cyc_ens} 
\end{table}
\vspace{-0.2cm}
\begin{table}[h]
	\caption{Respiratory disease detection - Task 2} 
        	\vspace{-0.2cm}
    \centering
    \scalebox{0.9}{
    \begin{tabular}{c c c c c c c} 
        \hline 
	    \textbf{Task}  &\textbf{Systems}    &\textbf{Spec.}   &\textbf{Sen.}   &\textbf{AS/HS Scores}  \\
        \hline 
	    Task 2                 &Baseline            &0.59             &0.75            &0.67/0.66  \\
	    Task 2                 &Two Scal.               &\textbf{0.65}             &\textbf{0.79}            &\textbf{0.72/071}  \\
	    Task 2                 &Gam. \& Scal.      &\textbf{0.65}             &0.76            &0.70/0.70  \\
       \hline 
    \end{tabular}
    }
    \label{table:ent_ens|} 
\end{table}
\vspace{-0.2cm}
Experimental results in Tables \ref{table:cyc_ens} and \ref{table:ent_ens|} show that ensemble of multiple spectrograms helps to improve the performance compared to the baseline. While ensemble of Scalogram and Grammatonegram achieves the best scores of 0.51/0.43 in Task 1, Task 2 shows the highest scores of 0.72/0.71 from ensemble of two Scalograms.

\subsection{Performance Comparison to the state of the art}

Given the results of inception-based networks and multiple spectrogram ensembles, we combine \textit{inception-01} network architecture and ensemble of Scalogram \& Gammatonegram for further analysis and compare the obtained results with the state-of-the-art systems (note that we only compare with systems that follow the standard ICBHI splitting of 60/40 with respect to subject independency~\cite{icb_ratio}).

As shown by the comparison in Table \ref{table:comp_cyc}, we achieve the scores of 0.53/0.45 in Task 1 that are very competitive to the state-of-the-art systems. Task 2 results presented in Table \ref{table:comp_ent} show that the ensemble system outperforms the state-of-the-art systems, but not better than the standalone system using \textit{inception-02} (cf. Table \ref{table:ent_inc}). 

\begin{table}[h]
    \caption{Comparison against state-of-the-art systems with ICBHI challenge splitting - Task 1 (highest scores in \textbf{bold}).} 
        	\vspace{-0.2cm}
    \centering
    \scalebox{0.9}{

    \begin{tabular}{l l l c c c} 
        \hline 
	    \textbf{Task}   &\textbf{Method}                        &\textbf{Spec.}   &\textbf{Sen.}   &\textbf{AS/HS Scores}  \\
        \hline 
        Task 1      &DT~\cite{ic_baseline}                    &0.75             &0.12           &0.43/0.15  \\        
        Task 1      &HMM~\cite{ic_hmm_18_sp}                   &0.38             &\textbf{0.41}           &0.39/0.23  \\        
        Task 1      &SVM~\cite{ic_svm_18_sp}                   &0.78             &0.20           &0.47/0.24  \\
        Task 1      &BRN~\cite{ma2019lung}                   &0.69             &0.31           &0.0.50/0.43  \\
        Task 1     &CNN-RNN~\cite{ic_cnn_19_iccas}          &\textbf{0.81} &0.28 & \textbf{0.54/0.42}    \\ 
        Task 1     &\textbf{Our system}                                &0.73    &0.32   &\textbf{0.53/0.45}    \\
       \hline 
    \end{tabular}
    }
    \label{table:comp_cyc} 
            	\vspace{-0.5cm}
\end{table}
\begin{table}[h]
    \caption{Comparison against state-of-the-art systems with ICBHI challenge splitting - Task 2 (highest scores in \textbf{bold}).} 
        	\vspace{-0.2cm}
    \centering
    \scalebox{0.9}{

    \begin{tabular}{l l l c c c} 
        \hline 
	    \textbf{Task}   &\textbf{Method}                        &\textbf{Spec.}   &\textbf{Sen.}   &\textbf{AS/HS Scores}  \\
        \hline 
        Task 2      &CRNN~\cite{ic_cnn_20_ieee_bs}                    &-             &-           &0.72/-  \\        
        Task 2      &CNN-MoE~\cite{pham2020cnnmoe}                   &0.71             &0.98           &0.84/0.82  \\        
        Task 2     &\textbf{Our system}                                &0.88    &0.85   &\textbf{0.86/0.86}    \\
       \hline 
    \end{tabular}
    }
    \label{table:comp_ent} 
\end{table}
%
\section{Conclusion}
This paper has presented an exploration of inception-based deep learning models and an ensemble of multiple input spectrograms for detecting respiratory anomaly and lung diseases from auditory recordings.
By conducting extensive experiments on the ICBHI meta-dataset, we show that our best model, which uses \textit{inception-01} architecture and ensemble of Gammatonegram \& Scalogram, outperforms the state-of-the-art systems on both Task 1 and Task 2, thus validate the efficacy of deep learning for early diagnosis of respiratory diseases.

\addtolength{\textheight}{-12cm}   

\bibliographystyle{IEEEbib}
\bibliography{refs}

\end{document}